%% file: manuscript.tex
\documentclass[useAMS,usenatbib,longnamesfirst,usedcolumn]{mn2e} 
\usepackage{epsfig}
\usepackage{times}
\usepackage{subfigure}  
\title[Mass composition and distribution]{The Birmingham-CfA 
cluster scaling project - II. Mass composition and distribution}
\author[A. J. R. Sanderson and T. J. Ponman]
       {A. J. R. Sanderson$^{1,2}$\thanks{E-mail: ajrs@astro.uiuc.edu} and 
          T. J. Ponman$^{1}$, \\
 $^{1}$School of Physics and Astronomy, University of
        Birmingham, Edgbaston, Birmingham B15 2TT, UK\\
 $^{2}$Department of Astronomy, University of Illinois, 
        1002 West Green Street, Urbana, IL 61801, USA
       \\}
 \date{Accepted 2003 July 23.
      Received 2003 July 22;
      in original form 2003 March 11}

\pagerange{\pageref{firstpage}--\pageref{lastpage}}
\pubyear{2003}

\shortcites{avi99, bal01, ben00, bennett03, blu84, bog89, car96, ebe96, ebe98, 
 fre99, gir00, gir02, gir95, hoe01, hra00, kau99, kor98, moh96, mua01, 
 odrpack_ug, ome01, piz97, pon96, rin02, san03, sat00, som01, squ96, whi93}

\defcitealias{san03}{Paper~I}
\defcitealias{pon02}{Paper~III}
\defcitealias{san03c}{Paper~IV}

\newcommand{\rmsub}[2]{\ensuremath{#1_{\mathrm{#2}}}} 
\newcommand{\srel}[2]{\mbox{\ensuremath{#1 - #2}}} 
\newcommand{\Bj}{\rmsub{B}{j}}
\newcommand{\Bjsun}{\rmsub{B}{\sun,j}}
\newcommand{\cf}{{\textrm c.f.}}
\newcommand{\Chandra}{\emph{Chandra}}
\newcommand{\chisq}{\ensuremath{\chi^2}}

\newcommand{\eg}{{\textrm e.g.}}

\newcommand{\fbary}{\rmsub{f}{b}}
\newcommand{\fgas}{\rmsub{f}{gas}}
\def\h70{\rmsub{h}{70}} 
\newcommand{\ie}{{\textrm i.e.}}
\newcommand{\keV}{\ensuremath{\mbox{~keV}}}
\newcommand{\km}{\ensuremath{\mbox{~km}}}
\newcommand{\kmpspMpc}{\ensuremath{\km \ps \pMpc\,}}

\newcommand{\LB}{\rmsub{L}{B}}
\newcommand{\LBj}{\rmsub{L}{B,j}}
\newcommand{\LBjM}{\srel{M}{\TX}}
\newcommand{\LBjT}{\srel{\LBj}{\TX}}

\newcommand{\MLR}{\ensuremath{M/L}}
\newcommand{\MLRBjsun}{\ensuremath{(M/\LBj)_{\sun}}}
\newcommand{\MLRsun}{\ensuremath{(M/L)_{\sun}}}
\newcommand{\Mpc}{\ensuremath{\mbox{~Mpc}}}
\newcommand{\Msol}{\rmsub{M}{\odot}}
\newcommand{\MT}{\srel{M}{\TX}}

\newcommand{\pMpc}{\ensuremath{\Mpc^{-1}}}
\newcommand{\ps}{\ensuremath{\s^{-1}}}
\def\R200{\rmsub{R}{200}} 
\newcommand{\rc}{\rmsub{r}{c}}
\newcommand{\RV}{\rmsub{R}{v}}
\newcommand{\s}{\ensuremath{\mbox{~s}}}

\newcommand{\TX}{\rmsub{T}{X}}
\newcommand{\WMAP}{\textit{WMAP}}
\newcommand{\XMM}{\emph{XMM-Newton}}

\begin{document}

\maketitle

\label{firstpage}

\begin{abstract}
  
 \noindent
 
 We investigate the spatial distribution of the baryonic and non-baryonic
 mass components in a sample of 66 virialized systems. We have used X-ray
 measurements to determine the deprojected temperature and density
 structure of the intergalactic medium and have employed these to map the
 underlying gravitational potential. In addition, we have measured the
 deprojected spatial distribution of galaxy luminosity for a subset of this
 sample, spanning over 2 decades in mass. With this combined X-ray/optical
 study we examine the scaling properties of the baryons and address the
 issue of mass-to-light (\MLR) ratio in groups and clusters of galaxies.
  
 We measure a median mass-to-light ratio of \hbox{224\,\h70} \MLRsun\ in
 the rest frame \Bj\ band, in good agreement with other measurements based
 on X-ray determined masses. There is no trend in \MLR\ with X-ray
 temperature and no significant trend for mass to increase
 faster than luminosity: \mbox{$M \propto \LBj^{1.08 \pm 0.12}$}. This
 implied lack of significant variation in star formation efficiency
 suggests that gas cooling cannot be greatly enhanced in groups, unless it
 drops out to form baryonic dark matter.  Correspondingly, our results
 indicate that non-gravitational heating must have played a significant
 role in establishing the observed departure from self-similarity in low
 mass systems.  The median baryon fraction for our sample is
 $0.162\,\h70^{-3/2}$, which allows us to place an upper limit on the
 cosmological matter density, $\rmsub{\Omega}{m} \leq 0.27\,\h70^{-1}$, in
 good agreement with the latest results from \WMAP.
 
 We find evidence of a systematic trend towards higher central density
 concentration in the coolest haloes, indicative of an early formation
 epoch and consistent with hierarchical formation models.
 
\end{abstract}



\section{Introduction}
\label{sec:intro}
As the material most amenable to detailed study, the properties of
optically luminous matter have long been used to infer the distribution of
mass in the Universe. Clusters of galaxies provide an excellent laboratory
for this purpose, since they are sufficiently large that their global
properties reflect those of the Universe as a whole. 

Previous work has shown that both optical light and gas, in particular,
are more spatially extended than the dark matter within virialized systems
\citep{dav95}. Evidence of a monotonic rise in gas fraction with radius
has also been found across a wide range of mass scales in a very large
sample of virialized systems \citep[hereafter \citetalias{san03}]{san03},
and also in the recent analysis of \citet{cas03}, which is in broad
agreement with numerical simulations \citep{eke98,fre99}. This behaviour
has been further confirmed in a high quality \XMM\ observation of a
relaxed cluster \citep{pra02}, where the X-ray halo has been traced out
almost to the virial radius, although \Chandra\ observations of extremely
relaxed lensing clusters have revealed a flat gas fraction profile in a
small number of cases \citep{all02}. However, the latter observations only
probe the innermost $\sim$1/3 of the halo: our data, though of poorer
quality, extend beyond this region in many cases, and we benefit from
averaging over a large ensemble of virialized systems.

While the hot gas has been shown to depart systematically from
self-similarity under the influence of non-gravitational physics, the same
cannot be said of the dark matter, which is not directly affected by such
processes.  N-body simulations indicate that the dark matter of virialized
haloes should follow a universal profile, across a wide range of scale
sizes \citep{nav95}, apart from a mild trend in its concentration with mass
\citep{nav97,sal98,avi99,jin00b}.

Although the contribution to the baryonic mass component from the
individual galaxies in virialized systems is significantly less than that
from the X-ray emitting intergalactic medium, the galaxy component is of
great importance, none the less. As the ultimate endpoint of gas cooling,
as well as the source of non-gravitational energy injection into the
intergalactic medium (IGM), the behaviour of the stars is closely linked to
the properties of the gas.  In addition, stellar sources are responsible
for the synthesis of the metals that contaminate the IGM, which are
released via supernova-driven winds \citep[\eg][]{fin01b}.

However, the scaling properties of the stellar content are rather less well
established than those of the IGM. In particular, there is some debate on
the mass-to-light ratio in groups and clusters and how this compares with 
the global value. Recent studies have reached differing conclusions about
how this quantity scales with halo mass. For example, \citet{gir02} report 
a trend towards increased \MLR\ in clusters, a result also favoured by 
\citet{mar02}. By contrast, measurements of \MLR\ based on X-ray mass 
estimates have concluded that it is roughly universal in groups and clusters
\citep{cir97,hra00}.

The approach of a combined X-ray and optical analysis of clusters and
groups of galaxies is a powerful tool for the understanding of mass
distribution in these systems. The X-ray data yield information about the
behaviour of the diffuse, hot intergalactic medium (IGM) and can be used to
determine the underlying gravitational potential structure, under the
assumption of hydrostatic equilibrium. By contrast, optical measurements
can be used to map the spatial properties of the baryons resident in
luminous matter, and hence deduce their contribution to the total mass
budget.

In \citetalias{san03}, we presented a detailed X-ray analysis of the IGM in
a large sample of 66 virialized systems, allowing us to reconstruct the
properties of the IGM. We have assembled our sample from three existing
X-ray studies of groups and clusters, to which we have added a small number
of cool groups. To each system, we have fitted analytical profiles to
parametrize both the gas density and temperature as a function of radius.
This allows us to place all the X-ray data on a unified footing, giving us
the freedom to extrapolate the gas properties to arbitrary radius. As with
detailed X-ray analyses, previous combined X-ray/optical studies of this
nature have been restricted to relatively small sample sizes
\citep[\eg][]{cir97,hra00}. In this paper, we build on our modelling of the
IGM, by incorporating the spatial distribution of stellar material, which
also allows us to determine the distribution of dark matter. Despite only
covering half our original sample (some 32 groups and clusters), our
optical sample is well suited to the study of the scaling properties of
these systems, since we retain good coverage across a wide range of system
masses.

Throughout this paper we adopt the following cosmological parameters;
$H_{0}=70$\kmpspMpc and $q_{0}=0$. All quoted errors are $1 \sigma$ on 
one parameter, unless otherwise stated.

\input{table1} 

\section{3D Galaxy Density Calculation}
\label{sec:opt_3Dfit}
The task of mapping the stellar mass in virialized systems is complicated
by the discrete nature of the individual galaxies -- unlike the evenly
distributed gas, which smoothly traces the underlying gravitational
potential. In addition, the contrast against the background of a cluster or
group of galaxies is much lower at optical wavelengths, scaling only
linearly with number density, rather than in proportion to
{\rmsub{\rho}{gas}}$^{2}$ in the case of X-ray emission. These two
difficulties are further compounded by the large angular size subtended by
a typical group or cluster, which may exceed one degree on the sky --
significantly larger than can be viewed by many current generation
wide-field CCD cameras.  Correspondingly, there have been comparatively few
detailed photometric studies of rich clusters made with CCDs
\citep[\eg][]{car97}.

One way to avoid some of the above problems is to quantitatively establish
cluster membership with redshift measurements of individual galaxies in its
vicinity. Although this avoids the need to estimate the background
contribution, such an advantage comes only at the price of long observing
times, in order to obtain high-quality spectra for a large fraction of the
galaxy members. For this reason such studies are restricted to a small
number of clusters \citep[\eg][]{kor02,kor00,moh96,gir95,fab89}. Given
these limitations, we have taken a different approach to the issue of
measuring the spatial distribution of galaxies and then estimating the
total luminosity of the whole system. We outline below our method, which is
based on widely-available digitized photographic plate data derived from
all-sky surveys.

The process can be separated into two stages. Firstly, determination of the
spatial distribution of the galaxy number density (see
section~\ref{sec:SDfit}) and, secondly, the calculation of the
normalization of this density profile (see section~\ref{ssec:lum_norm}).
Although the normalization can, of course, be calculated in the spatial
fitting, this makes no allowance for the contribution to the total
luminosity from galaxies too faint to be observed. To estimate this
contribution it is necessary to determine the luminosity function of the
observed galaxies and extrapolate this down to some limiting magnitude, in
order to correct for the missing light. Whilst it is possible to do this
using photographic plate based data, the measured magnitudes require
careful calibration, due to subtle variations in the sensitivity of the
photographic emulsion both within and between plates.  Therefore, we have
taken integrated luminosity values from a small number of sources in the
literature and used these to infer a 3-dimensional normalization, in
combination with our fitted profile parameters.

We assume a distribution of optical light described by an NFW profile 
\citep{nav95}, \ie
\begin{equation}
\rho = \frac{\rho_0}{x \left( 1 + x \right)^2 },
\label{eqn:NFW}
\end{equation}
where $x=r/\rmsub{r}{s}$ and \rmsub{r}{s} is a characteristic scale radius.
This function rises from a logarithmic slope of -3 at large radii to a
central cusp ($\rho \propto r^{-1}$) with the transition between the two
regimes occurring around \rmsub{r}{s}. Such a parametrization is
advantageous as it has only one free parameter (since the normalization is
constrained by our maximum likelihood fitting method -- see
section~\ref{sec:SDfit}), which helps stabilize the fitting, given the
relatively large background level in the case of the plate data.

We make the assumption of spherical symmetry in the deprojection of the
optical light distribution, as we did for the X-ray analysis in
\citetalias{san03}. Since our sample has been selected on the basis of a
relaxed X-ray morphology, it is reasonable to expect a corresponding degree
of regularity in the galaxy distribution. This also improves the quality of
the fit, which might otherwise be degraded by the presence of significant
substructure or bi-modality. However, despite this, it was not possible to
obtain a satisfactory surface density fit to the photographic plate data
for a small number of cool groups; their treatment is described in
section~\ref{ssec:NED_fit} below.

\subsection{APM data}
\label{ssec:APM_fit}
The Automatic Plate Measuring (APM) machine source
catalogue\footnote{http://www.ast.cam.ac.uk/\~{}apmcat/} is a digitized
catalogue based on the 1st generation Palomar Observatory Sky Survey
(POSS~I) and United Kingdom Schmidt Telescope (UKST) photographic plates.
The field of view covered is extremely large -- 6.2 and 5.8 degrees on a
side for the plates from POSS~I and UKST, respectively -- and the whole sky
has been observed in this way. As such, it is ideal for studying large scale structure at
optical wavelengths, although the magnitude information it provides can be
rather unreliable, owing to saturation and other problems with photographic
emulsions.  However, we only exploit the positional data available -- which
are very reliable -- since we determine total luminosities for our groups
and clusters from literature measurements.

To extract source positions for our fitting, we have applied very
conservative selection criteria to allow for problems with source
identification. Although detection algorithms have been used to distinguish
between stellar and non-stellar sources, this classification is often
unreliable and can cause stars to be identified as galaxies
\citep[\eg][]{car00} and vice versa: in a recent analysis of data from the
Minnesota Automated Plate Scanner (APS) digitized survey of POSS~I plates,
\citet{rin02} found that $\sim$3.3 per cent of objects identified
as stars were actually galaxies, in the vicinity of the cluster Abell~2199.
This confusion is exacerbated for faint galaxies, and consequently we only
neglect sources classified as stars if they are brighter than $\rmsub{m}{b}
= 15$, otherwise all detected sources are used. This avoids any potential
systematic bias in the fitting between clusters, at the expense of a
significantly increased background level. All objects classified as noise
features are excluded.

\subsection{NED data}
\label{ssec:NED_fit}
The problems associated with optical analysis are exacerbated at the scale
of groups of galaxies, which present even lower contrast against the
background, owing to their smaller volume. However, many well-studied
groups -- including most of those in our sample -- are sufficiently close
so as to have measured recession velocities available in the literature. In
such cases it is therefore possible to establish unambiguous membership of
the group, thus eliminating the contamination from fore- and background
objects. However, the numbers of such confirmed members are generally
rather small and tend to be predominantly limited to the inner regions of
the halo. The impact this has is mitigated to some extent by our chosen
parametrization of galaxy density (equation~\ref{eqn:NFW}), which has only
a single free parameter (the scale radius, \rmsub{r}{s}), since sparse data
obviously provide weaker constraints on spatial fitting
\citep[\eg][]{gir95}.

It was not possible to derive NFW scale radii from the APM data for four
of our groups (NGC~2563, NGC~5129, NGC~5846 \& NGC~6338) and so we took
galaxy positions from the NASA Extragalactic Database (NED) for these
systems.  We used only those galaxies which were identified as being group
members in the database. It was also necessary to do this for Abell~539,
since no POSS~I or UKST data were available for this poor cluster. There
is potential for bias when using NED data, since the completeness
properties of the database are unknown -- although we are only concerned
with the spatial completeness, since the luminosities for these systems
are taken from other sources. In particular, it is possible that there may
be a dearth of galaxies catalogued in the outer regions of these groups,
which may introduce a spurious central surface density enhancement.
However, we note that any such bias is reduced by the fact that we have
used the radius of the outermost galaxy as the cut-off radius in the
fitting, rather than the X-ray determined virial radius used to select the
galaxies from NED.

\subsection{Surface Density Fitting}
\label{sec:SDfit}
As we are interested in the 3-dimensional properties of the galaxy optical
light, we have chosen to parametrize the space density distribution in
three dimensions and then numerically project this, for comparison with the
surface density of galaxies. The same approach was used in the deprojection
of the X-ray data \citepalias[see][]{san03}. This method is advantageous,
since it directly yields the required parameters, without the need for
inversion.  For the analysis of the APM data, a constant background term
was included as a free parameter, to allow for a variation in the number of
fore- or background sources caused, for example, by a differences in
large-scale structure along the line-of-sight. It is important to fit this
separately for each cluster, since our background (incorporating a stellar
component) can vary significantly with position on the sky. The centroid of
the model was fixed at the cluster position as listed in NED, in order to
stabilize the fitting. Although the optical centroid can be displaced with
respect to the X-ray centroid, we note that \citet{gir00} find a typical
error in \rmsub{L}{B} within \RV\ of just 5 per cent, when recomputing
their integrated optical luminosities using an X-ray rather than optical
centroid.

The 3-dimensional density profile was evaluated in a series of spherical
shells, by numerically integrating equation~\ref{eqn:NFW} between their
radial boundaries. For each spherical shell there is a corresponding
annulus, with identical radial bounds. The surface density in each annulus
is calculated by summing the contributions made to it by its corresponding
shell and all those lying outside it. Hence the contribution to the surface
density from projection along the line of sight is fully accounted for.

We have chosen to perform an \textit{unweighted} fit to the data (as also
used by \citealt{hra00} and \citealt{kor02}, for example), so that each
galaxy is treated as a single point at a given position. This approach is
advantageous since it avoids the need to use plate-measured magnitudes,
which require careful calibration as mentioned above. By treating each
galaxy identically we are implicitly assuming that there is no luminosity
segregation, \ie\ that the luminosity function is everywhere the same
within the cluster. This is a reasonable assumption, since any segregation
is generally limited to only the very brightest galaxies -- which tend to
be located more centrally in clusters \citep{ada98}. 

The use of an unweighted fit has important consequences for those clusters
that possess a central cD galaxy. On the one hand, our method limits the
potential for an extremely bright, centrally located galaxy to skew the
fit, by overemphasising the central luminosity density. On the other hand,
in so doing, there is a tendency to slightly overestimate the luminosity in
the outer regions. This is due to our method for determining the luminosity
normalization, which does include a contribution from any central cD
object. As a result, the extra luminosity of such a galaxy is, in effect,
`spread out' over the whole profile, leading to an artificial excess in the
outer regions. However, the effect this has on our total luminosity
estimates is small, given the generally large sizes of the apertures used
to measure the luminosity values to which we normalize our density profiles
(see section~\ref{ssec:lum_norm}). In any case, the contribution to the
luminosity function from cD galaxies is fully accounted for by this same
method for determining the total cluster luminosity.

Although some previous studies of galaxy distributions in clusters have
fitted analytical profiles with cores to the data
\citep[\eg][]{gir95,cir97}, we favour the cusped NFW profile
(equation~\ref{eqn:NFW}). Not only does it have one fewer fitted parameter
-- which helps stabilize the fitting, as previously mentioned -- it also
provides a more accurate description of the dark matter distribution
\citep{nav95}. Simulations indicate that the galaxies trace the dark matter
more closely than the gas does, with only a modest bias \citep{dave02};
they may even be more centrally concentrated, owing to the effects of
dynamical friction \citep{met97}. We note that \citet{ada98} have shown
that only the surface density of faint galaxies shows a significant
preference for a distribution with a core rather than a cusp.  Since our
analysis is based on photographic plate observations, rather than the
higher quality CCD data used in their work, we can therefore expect to be
relatively insensitive to this effect.

Previous studies offer support for the presence of a cusp in the luminosity
distribution. For example, \citet{oeg87} measure a power law slope of
approximately -1 in the centres of 3 Abell clusters; \citet{mer94} find
evidence of a central power law dependence of galaxy space density in the
core of the Coma cluster, using a non-parametric algorithm. \Citet{bee86}
have demonstrated that imperfect centring can `erase' the presence of a
real cusp and artificially introduce a constant density core. Using two
methods of determining the cluster centre, they find that the majority of
the 48 clusters in their sample posses a central cusp in projected galaxy
density. The inner regions of these clusters are well characterized by a
power law slope of -1.

We use the maximum likelihood fitting procedure described in \citet{sar80},
which we now briefly outline. There are three parameters to be determined
-- the scale radius, \rmsub{r}{s}, the surface density normalization,
$\Sigma_0$, and background surface density, \rmsub{\Sigma}{bg} (since the
centroid is fixed -- see above). However, only two of these parameters are
independent, therefore we require an additional constraint, otherwise the
maximum likelihood solution is unbounded. We apply the condition that the
model reproduce the observed number of galaxies within the fitting region.
In the standard method of \citet{sar80}, \rmsub{r}{s} and $\Sigma_0$ are
left as free parameters and \rmsub{\Sigma}{bg} is calculated so that the
`area' under the model + background equals the total number of galaxies
within the fit region. However, in the case of our NED data
\rmsub{\Sigma}{bg} is known a priori (\ie\ it is zero, since all the
galaxies are confirmed group members), so we use the alternative method
detailed in section~IV of \citet{sar80}. Here, \rmsub{\Sigma}{bg} and
\rmsub{r}{s} are left as fitted parameters, while $\Sigma_0$ is instead
calculated so as to reproduce the observed galaxy count.

As a result of our numerical projection technique, $\Sigma_0$ is not
actually a direct model parameter. However it is trivially related to
$\rho_0$, the NFW central density normalization from
equation~\ref{eqn:NFW}. We determine the appropriate value of $\rho_0$ by
setting it initially to unity and calculating the `area' under the
projected data, excluding the parts of any annuli lying outside the fit
region, \ie\ summing the product of the surface density in each annulus and
its area. The ratio of the observed number of galaxies in the fit region to
this number then yields $\rho_0$.

Fitting was performed using the gradient-based MIGRAD method in the
\textsc{minuit} minimization library from CERN \citep{minuit_ug} and
errors on the parameters were found with MINOS, from the same package.
Errors were determined from the increase in the Cash statistic of one,
since differences between values obtained from the same data set are
\chisq-distributed.

\section{Optical Luminosity Calculation}
\label{sec:opt_lum}
Since we have chosen to avoid using magnitude information in the
calculation of the galaxy surface density distribution, we require an
alternative means of deriving a normalization value for our 3-dimensional
luminosity profiles.  To do this we have taken integrated luminosity
measurements in fixed metric apertures from the literature, which allow 
us to infer a central luminosity density. The details of this procedure 
are described in section~\ref{ssec:lum_norm} below. 

The majority of these values are taken from the sample of \citet{gir02},
which incorporates most of the sample of \citet{gir00}. This sample
comprises data from the APS and COSMOS/UKST surveys -- we have selected
luminosity estimates from the latter for preference, where overlaps
occurred. The only exception to this is HCG~62, for which the COSMOS
aperture luminosity quoted in \citet{gir02} is a factor of 2 larger than
the APS luminosity; for this group we have chosen the APS measurement. We
note that the APS luminosity leads to a mass-to-light ratio
section~\ref{ssec:MLR_integrated}) which is more consistent with the
simple estimate of \citep{pon93}. Data for three other clusters (Abell~539,
Abell~2256 and AWM~7) were taken from \citet{gir00}. These latter systems
have two different luminosity estimates based on literature magnitudes --
we chose the `M(red)' sample, since this provided the largest overall
number of galaxies detected for all three clusters.

A further five systems are covered by the study of \citet{hra00}, based on
data from the second generation Palomar Observatory Sky Survey (POSS~II),
calibrated with CCD photometry. Two groups (NGC~2563 and NGC~5846) have
luminosity estimates from \citet{hel03} within an aperture defined by the
virial radius as calculated in \citet{hel00}. Finally, we used the
luminosity quoted in \citet{squ96} for the rich cluster Abell~2218. The
references for each system are listed in Table~\ref{tab:sample}, together
with some key properties.

\subsection{Conversion between bands}
\label{ssec:band_convert}
We have adopted the \Bj\ photometric band as our standard reference frame,
since this was used by \citet{gir02} and \citet{gir00}, and we take $\Bjsun
= 5.33$ \citep{gir02}.  However, our other literature sources have used
different bands and so we have converted these values into the \Bj\ band,
under the assumption that the majority of the light originates in
early-type galaxies. We have used the following relations to perform the 
conversions.

To convert from the $B$ band to the \Bj\ band, we assume 
\begin{description}
\item $B_{\sun} = 5.48$ \citep{gir00}
\item $\Bj = B - 0.28(B-V)$ \citep{bla82}
\item $B-V = 0.9$, for early-type galaxies \citep{gir00}
\item $\Rightarrow \Bj = B - 0.252$,
\end{description}
which leads to
\begin{equation}
\label{eqn:B2Bj}
\frac{ L_{\Bj} }{ L_{\Bjsun} } = 1.10 \frac{ L_{\textrm{B}} }{ L_{\textrm{B},\sun} }.
\end{equation}

To convert from the $V$ band to the \Bj\ band, we assume
\begin{description}
\item $V_{\sun} = 4.82$ \citep{all73}
\item $\Rightarrow \Bj = V + 1.152$, from above (for early-type galaxies)
\end{description}
which leads to
\begin{equation}
\frac{ L_{\Bj} }{ L_{\Bjsun} } = 0.554 \frac{ L_{\textrm{V}} }{ L_{\textrm{V},\sun} }.
\end{equation}

For comparison with $R$ band luminosities, we assume
\begin{description}
\item $R_{\sun} = 4.28$ \citep{all73}
\item $V-R = 0.55$ for early-type galaxies \citep{gir00}
\item $\Rightarrow \Bj = R + 1.702$, from above,
\end{description}
which leads to
\begin{equation}
\frac{ L_{\Bj} }{ L_{\Bjsun} } = 0.547 \frac{ L_{\textsc{R}} }{ L_{\textsc{R},\sun} }.
\end{equation}

\subsection{Determination of luminosity normalization}
\label{ssec:lum_norm}
With the exception of the \citet{hra00} data, all our literature values of
luminosity are simple estimates within a fixed aperture on the sky. As
such, these measurements represent the integrated luminosity within a
cylinder, defined by the aperture multiplied by the diameter of the
cluster.  It was therefore necessary to allow for the projected
contribution to this value from luminosity at large radii, when inferring
values for the 3-dimensional normalization.  We accomplished this in the
following way, using our numerical projection technique described
previously.  The normalization was set to unity and the NFW profile was
integrated out to a large radius, to insure convergence (taken to be 500
arcmin), in a series of fixed-width spherical shells. The projected
emission from these shells was then computed, in the corresponding series
of annuli. The total luminosity within the aperture radius was then
calculated, by simply summing the contributions from the annuli within that
radius. Where the aperture radius lay within an annulus, the contribution
from this annulus was determined by linear interpolation, since the annulus
width is small. The appropriate normalization factor was found by dividing
the measured luminosity value by that calculated for a normalization of
unity.

For the five clusters analysed by \citet{hra00}, a slightly different
approach was taken, since the authors corrected their aperture luminosities
for the effects of projection. Working in three dimensions, we set our
normalization to unity and simply integrated our density profile from $r=0$
to a radius of 1h$_{50}$\Mpc\ (used for all their quoted luminosities). The
ratio of the aperture luminosity to the integrated value then yields the
central luminosity density that we require.

The errors on our quoted luminosities are derived from the errors on the
aperture values. Following \citet{gir02}, we assume a 1$\sigma$ error of 50
per cent on our total \LBj\ values based on their data. The same fractional
error was assumed for the values from \citet{hel03}, based on their own
error estimates. The values for the \citet{hra00} data have smaller errors,
as does the \citet{squ96} luminosity for Abell~2218, since these are based
on higher quality POSS~II and CCD observations, respectively, which are
more accurate than the majority of our photographic plate data.

Owing to our choice of an NFW profile for the fitting, coupled with the
bias towards optically rather than X-ray selected clusters in previous
studies, we are extremely restricted in the number of direct comparisons
which can be made between our spatial fitting results and those available
in the literature.  However, we note that \citet{ada98} found a scale
radius of $3.7\pm0.6$ arcmin for the cluster Abell~119, which is reasonably
consistent with our value of 5.5 arcmin (see Table~\ref{tab:sample}). This
change in $r_s$ would give a $\sim$4 per cent change in the total optical
luminosity within \R200, based on the aperture luminosity for this cluster
as measured by \citet{gir02}. The impact of uncertainties in the spatial
fitting parameters is limited by the fact that the apertures from which the
optical luminosities we use for calibration have been taken, are close to 
\R200: the median difference between these aperture radii and our X-ray 
determined values of \R200\ is $\sim$50 per cent. In the case of Abell~119,
our correction to the \citet{gir02} aperture luminosity amounts to only
$\sim$10 per cent.  For Abell~2256, the analysis of \citet{oeg87} provides
support for our findings, measuring an inner logarithmic slope of
$-0.98\pm0.02$ for the galaxy density distribution in this
cluster. Moreover, their binned radial profile is consistent with an outer
slope of -3 and a break radius similar to our value of 7.3 arcmin.

\section{Results: Optical Properties}
\label{sec:results:optical}
We convert optical light directly into stellar mass, assuming a
mass-to-light ratio for early-type galaxies of \rmsub{(M/L)}{B} =
7\,\h70 Solar \citep{piz97}. A similar value of ($8.3\pm0.35$)\,\h70\ 
was found by \citet{van91}. Using \mbox{$B_{\sun} - \Bjsun = 0.15$}, from
section~\ref{ssec:band_convert} gives \mbox{$\rmsub{(M/L)}{\Bj} =
  6.1\,\h70$\MLRBjsun}.  To infer the density profile of dark matter, we
subtract the gas and stellar mass from the total mass (as determined from
the X-ray data).  Consequently, we only have information on the dark matter
and stellar properties for the optical sub-sample of 32 groups and
clusters.

Relationships between integrated properties provide an important tool for
investigating the similarity between haloes across a range of masses.
Although quantities such as luminosity are directly observable, some
allowance must be made for the effects of projection if fair comparisons
are to be made between different systems. Correspondingly, all our quoted
luminosity values (see Table~\ref{tab:sample}) are derived from 
integrating the \emph{3-dimensional} light profiles, using the appropriate
normalization calculated in section~\ref{ssec:lum_norm}, to provide a more
sensitive probe of scaling properties. Where we have compared our results
with those from the literature quoted in different photometric bands, we
apply the correction factors given in section~\ref{ssec:band_convert} to
determine equivalent values in the \Bj\ band.

\subsection{Optical luminosity}
\begin{figure}
\hspace{-0.3cm}
\includegraphics[angle=270,width=9cm]{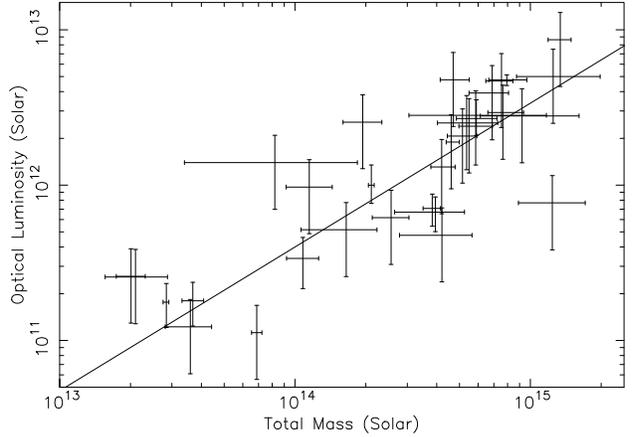}
\caption{ \label{fig:MTOT-LBj}
  Total gravitating mass as a function of \LBj\ luminosity (both within
  \R200), for the optical sample. The line shows the best-fitting
  power-law, which has a logarithmic slope of $0.96 \pm 0.10$.}
\end{figure}

A useful test of the scaling properties of the stellar distribution is the
relationship between optical luminosity and mass.  Fig.~\ref{fig:MTOT-LBj}
shows the \LBj\ luminosity within \R200\ for the optical sample plotted
against the total mass within \R200, as calculated in \citetalias{san03}.
We performed an orthogonal distance regression to fit a straight line to the
data in log space, using the \textsc{odrpack} software package
\citep{bog89,odrpack_ug}, to take account of errors in both X and Y
directions. The best-fitting relation is $\log{\LBj}=(-1.36\pm
1.46)+(0.96\pm0.10)\times\log{M}$ and the scatter in the data is 1.21 times
that expected from the statistical errors alone. The logarithmic slope of
this relation is flatter than, but consistent with self-similarity
(\ie\ $\LBj \propto M$), at the 1$\sigma$ level -- this reflects the slight
excess in stellar density in the coolest systems referred to in
section~\ref{ssec:lum(r)}. Performing the regression with the axes 
reversed leads to a scaling between light and mass of 
$M\propto \LBj^{1.08\pm 0.12}$.

\begin{figure}
\hspace{-0.3cm}
\includegraphics[angle=270,width=9cm]{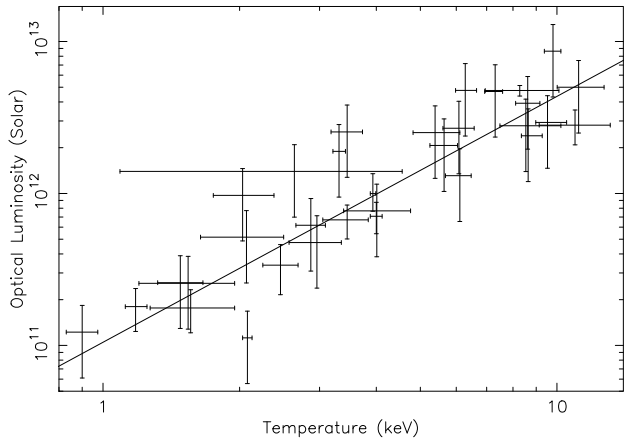}
\caption{ \label{fig:LBj-kT}
  Total \LBj\ luminosity within \R200\ as a function of system temperature
  for the optical sample. The line shows the best-fitting power-law, which 
  has a logarithmic slope of $1.62 \pm 0.14$.}
\end{figure}

For comparison, we have also plotted \LBj\ against the emission-weighted
temperature, as calculated in \citetalias{san03}, which we show in
Fig.~\ref{fig:LBj-kT}. Using the same regression technique, the
best-fitting relation is $\log{\LBj}=(11.02 \pm 0.01)+(1.62 \pm
0.14)\log{kT}$, and the scatter in the data is 0.96 times that expected
from the statistical errors alone. Furthermore, there is no evidence of any
steepening in the relation or any other systematic deviation from a simple
power law, indicating that this provides an accurate description of the
data. The logarithmic slope of this relation is steeper than, but
consistent with self-similarity (\ie\ $\LBj \propto T^{1.5}$, assuming that
light traces mass), at the 1$\sigma$ level. However, this is due to the
effect of the \MT\ relation slope -- which is itself significantly steeper
than the self-similar relation \citepalias{san03}.

The scatter about the \LBjT\ relation is rather less than that about the
\LBjM\ relation, especially considering that our mass errors are quite
conservative \citepalias{san03}. This is puzzling, but may reflect the fact
that the optical luminosity is dominated by the stellar contribution from
the inner regions, where the gas temperature is also more heavily weighted.
This would produce a tighter correlation between \LBj\ and $kT$ than
between \LBj\ and $M$, since the total mass is less sensitive to the
contribution from the inner regions of the halo.

\subsection{Mass-to-light ratio}
\label{ssec:MLR_integrated}
Since light is easily observed, it is often used as a tracer of
mass in the Universe, expressed as the ratio of total mass to optical
luminosity in some photometric band. Assuming that clusters of galaxies
are a fair representation of mass composition on large scales, the
mass-to-light ratios in these systems can be used to estimate the total
mass density of the Universe. It is therefore important to understand the 
scaling properties of this quantity if such cosmological inferences are 
to be unbiased.

\begin{figure}
\hspace{-0.3cm}
\includegraphics[angle=270,width=9cm]{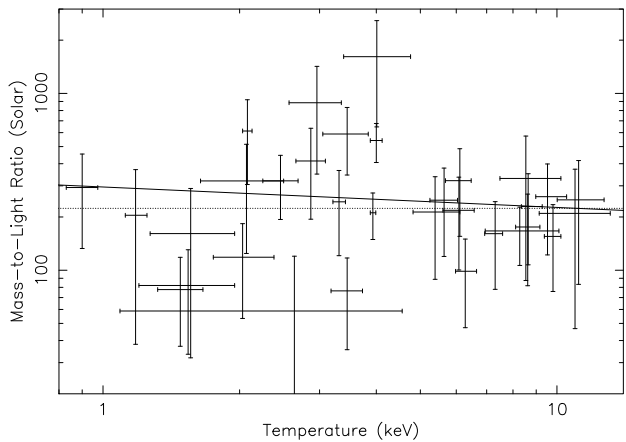}
\caption{ \label{fig:MLR_kT}
  Mass-to-light ratio within \R200 (in the \Bj\ band) as a function of
  system temperature for the optical sample. The solid line represents the
  best fitting power law, which has a logarithmic slope of $-0.11\pm0.18$.
  The dotted line shows the median value for the whole sample, of
  224\,\h70\ \MLRBjsun.}
\end{figure}

Fig.~\ref{fig:MLR_kT} shows the variation in mass-to-light ratio with X-ray
temperature, for the optical sample (see Table~\ref{tab:sample} for the
data). It can be seen that there is little evidence of any trend in the
data; Kendall's K statistic indicates the significance of a negative
correlation is $0.37\sigma$. The best-fitting power law has a logarithmic
slope of $-0.06\pm0.17$, which is fully consistent with no trend.  The
level of scatter about the best-fit is exactly that expected from the
statistical errors. We therefore conclude that a universal mass-to-light
ratio in groups and clusters provides a good description of most of our
data.

It can be seen that one point in particular is very high on this relation:
the poor cluster AWM~7 has $(M/\LBj) \sim 1600$\MLRBjsun\, albeit with
rather large errors. A large mass-to-light ratio was also measured for
this system by \citet{kor98}, who found a value of $(650 \pm
150)$\,\rmsub{h}{100} \MLRsun\ for the $R$ band, corresponding to $(830
\pm 190)\,\h70$ \MLRBjsun, consistent with our $1\sigma$ lower bound of
$\sim$650. Exclusion of this point lowers the mean of our sample from $308
\pm 53$ to $266 \pm 32$\,\h70 \MLRBjsun. We also note that the
mass-to-light ratio for HCG~62 is rather low -- only $78\pm40$, which is
rather less than the value estimated in \citet{pon93}. As was mentioned in
section~\ref{sec:opt_lum}, there is a factor of two difference between the
aperture luminosities quoted in \citet{gir02}, based on APS and COSMOS
data for this system. Furthermore, we note that the optically-determined
virial radius used for their aperture is twice as large as our X-ray
measured \R200, which means that we have to \emph{reduce} the
\citet{gir02} luminosity value, according to our fitted galaxy density
profile.

To minimize the bias caused by such discrepant measurements in our \MLR\ 
data, we have evaluated a \emph{logarithmic} mean value, of
$233^{+32}_{-28}$\,\h70\ \MLRBjsun. This is significantly lower than the
ordinary, arithmetic mean value of $308 \pm 53$, but compares very well
with the median value of 224\,\h70\ \MLRBjsun. This illustrates the effect
that anomalously large \MLR\ systems in particular can have, when
determining a representative average.

In a combined X-ray/optical study of 12 Abell clusters of galaxies,
\citet{cir97} measured a mean mass-to-light ratio equivalent to 346\,\h70
\MLRBjsun, which is consistent with our results. More recently,
\citet{hra00} have also determined \MLR\, for a sample of 8 groups and
clusters of galaxies, correcting their optical luminosities for projection
effects, by assuming a spatial distribution of light identical to that they
measured for the X-ray gas. Their mean value of \MLR\ is equivalent to $310
\pm 45$ \MLRsun\ in the \Bj\ band, which is in excellent agreement with our
own arithmetic mean. Furthermore, their median \MLR\ of $\sim$250 is
indistinguishable from our own median value.  \Citet{hra00} also conclude
that \MLR\ is roughly independent of mass, as was also found by
\citet{dav95}, albeit based on a rather small sample in both cases. We note
however, that their assumption that the optical light traces the gas mass
is not supported by our data (see section~\ref{ssec:MLR_radial}), although
this is likely to have only a small effect on their results.

The agreement with \MLR\ as measured using optical mass estimates is
generally less good, as pointed out by \citet{hra00}. For example,
\citet{gir00} quote a mean value of $\sim$175\,\h70 \MLRBjsun\ for their
sample of 105 clusters, from which many of our own luminosity estimates
have been taken (since many of their results were incorporated in
\citet{gir02}). However, the disagreement with our own mean value of $308
\pm 53$\,\h70\ \MLRBjsun\ can be attributed to a difference in luminosity
as well as mass: we have extended their quoted aperture \LBj\ values,
according to our fitted galaxy density profiles, and evaluated an
integrated luminosity within \R200\ as derived from our X-ray data.  On
lower mass scales, optically-derived \MLR\ measurements are also lower than
our own value. \Citet{ram97} find a mean approximately equivalent to $151
\pm 60$\h70\, \MLRBjsun\ (99\% confidence) for a very large sample of
groups in the Northern CfA Redshift Survey, using virial mass estimates.

The difference between X-ray and optically derived masses can also be seen
in the scaling properties of \MLR. While our results support a universal
mass-to-light ratio in groups and clusters -- consistent with the X-ray
studies of \citet{hra00} and \citet{cir97} -- \citet{gir02} find that mass
increases more quickly than luminosity, such that $M \propto \LB^{1.34 \pm
  0.03}$. However, the optical study of \citet{car96} indicates that \MLR\
is universal in clusters, although their mean value, equivalent to $380
\pm 70$\,\h70\ \MLRBjsun\ is somewhat higher than our own. Weak lensing
provides the most robust determination of halo mass, and the analysis of
\citet{hoe01} indicated that group mass-to-light ratios were lower than
those of clusters, although their sample comprised only groups. Such a
variation of \MLR\ with halo mass is confirmed by the work of
\citet{bah02}, who incorporated lensing mass estimates in part of their
sample.

Semi-analytical models (SAMs) of galaxy formation generally predict a
significant increase in \MLR\ with halo mass
\citep[\eg][]{kau99,ben00,som01}. As an example of the typical behaviour
observed, \citet{ben00} find that \MLR\ reaches a minimum on mass scales
of $\sim 10^{12}\Msol$, increasing by a factor of 3 up to halo masses of
$10^{15}\Msol$. This corresponds to a logarithmic slope of 0.16, which is
only just outside the 1$\sigma$ upper bound of 0.13 from our data in
Fig.~\ref{fig:MLR_kT}, although we only include groups and clusters in our
optical sample. This tendency for \MLR\ to increase away from a minimum on
the scale of a typical galaxy is also confirmed by \citet{mar02}.
Semi-analytical models also predict that \MLR\ in clusters is
significantly lower than the global value \citep{kau99}, suggesting that
even massive clusters are a biased estimator of the mean \MLR\ in the
Universe. That our findings do not provide a closer match to the \MLR\
predictions of these models may reflect the over-cooling of gas in SAMs,
which is a well-known problem that leads to the formation of excessively
bright central galaxies, for example \citep{kau93}.

\subsection{Stellar Distribution}
\label{ssec:lum(r)}
\begin{figure}
\hspace{-0.3cm}
\includegraphics[angle=270,width=9cm]{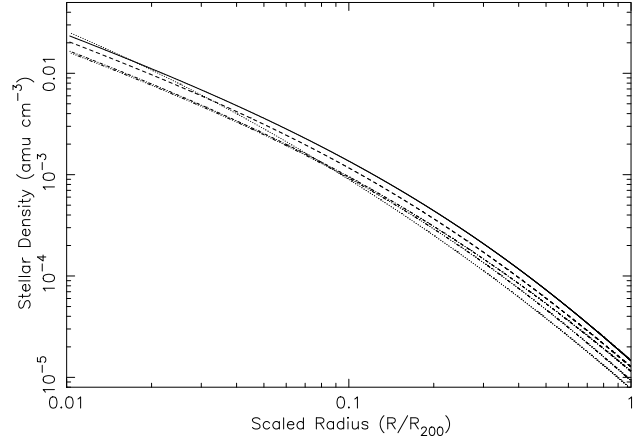}
\caption{ \label{fig:stellar_profiles}
  The variation of stellar matter density with scaled radius, for the
  \textbf{optical sample}, grouped by system temperature. The solid line
  represents the coolest systems (0.3--2.0\keV), increasing in temperature
  through dashed (2.0--2.9\keV), dotted (2.9--4.6\keV), dot-dashed
  (4.6--8.5\keV) and finally dot-dot-dot-dashed (8.5--17\keV).}
\end{figure}

The stellar mass density as a function of scaled radius, for the optical
sample, is shown in Fig.~\ref{fig:stellar_profiles}, grouped into five
temperature bins for clarity. There is a remarkably clear consistency in
the shape and normalization of the profiles across the whole range of
temperatures, which are approximately co-aligned.  However, close
inspection of Fig.~\ref{fig:stellar_profiles} reveals a slight excess in
the stellar density of the coolest two temperature bands, compared to the
more massive systems. This points towards a possible weak trend in star
formation efficiency with mass; we will revisit this issue in
section~\ref{ssec:SFE}.

\subsection{Spatial variation of mass-to-light ratio}
\label{ssec:MLR_radial}
\begin{figure}
\hspace{-0.3cm}
\includegraphics[angle=270,width=9cm]{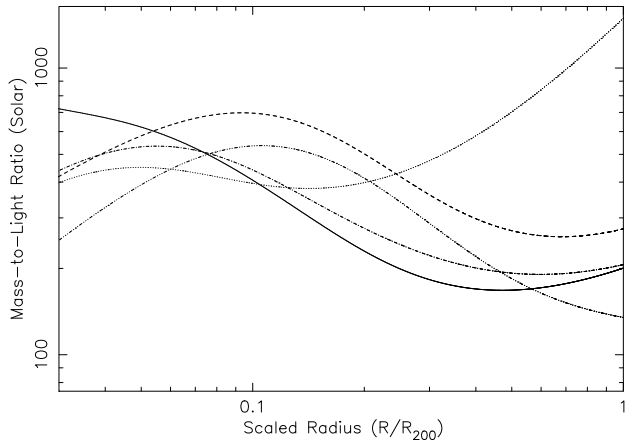}
\caption{ \label{fig:mlr_profiles}
  The variation of mass-to-light ratio (in the \Bj\ photometric band) with
  scaled radius for the \textbf{optical sample}, grouped by system
  temperature. The solid line represents the coolest systems
  (0.3--2.0\keV), increasing in temperature through dashed (2.0--2.9\keV),
  dotted (2.9--4.6\keV), dot-dashed (4.6--8.5\keV) and finally
  dot-dot-dot-dashed (8.5--17\keV).}
\end{figure}

The extent to which light traces mass \emph{within} virialized haloes can
be gauged by studying the mass-to-light ratio as a function of radius in
these systems. Fig.~\ref{fig:mlr_profiles} shows these profiles for the
optical sample, grouped by temperature as before. It can be seen that there
is evidence of decrease in this quantity with increasing radius, outside of
$\sim$0.1\R200, in four of the five bands. Interestingly, the central range
(2.9--4.6\keV) shows a clear \emph{increase} in \MLR\ with radius in this
region, although this may be partly driven by the anomalous behaviour of 
AWM~7, which exhibits a strongly rising mass-to-light profile, as also found
by \citet{kor98}.

Outside of $\sim$0.1\R200\, our results demonstrate that the distribution
of optical light is more extended than that of the gravitating mass, as
previously reported by \citet{dav95}. It is worth noting that the
simulations of \citet{met97} indicate that the stellar distribution can be
\emph{less} extended than that of the dark matter, due to the effects of
dynamical friction transferring energy away from the galaxies.

We have chosen to truncate the radial scaling in
Fig.~\ref{fig:mlr_profiles} at 0.03\R200, rather than 0.01\R200, as our
optical luminosities are likely to be unreliable in the core of the halo.
This is because our unweighted galaxy density fitting is insensitive to the
excess luminosity associated with a central cD galaxy. However, we would
therefore expect to \emph{underestimate} the luminosity in the core, which
would overestimate the mass-to-light ratio and hence flatten any drop in
\MLR\ in the centre. This suggests that the observed trend towards a
decrease in the mass-to-light ratio $\la0.1\R200$ is probably real. We also
note that a similar decline in \MLR\ towards the centre is seen in the
central regions of most of the 12 clusters in the sample of
\citet{cir97}. A recent detailed Chandra analysis of Abell~2029 also
reveals a sharp decrease in \MLR\ within 0.1\R200\ \citep{lew03}.

\section{Results: Dark Matter, Gas and Total Mass Properties}
\label{sec:results:non-optical}
Building on our analysis of the optical properties of our sub-sample of 32 
groups and clusters, we now present equivalent information for the other
mass components. Since we infer the dark matter distribution using our 
knowledge of the stellar mass profile, we are limited to our `optical 
sub-sample' in the analysis of its properties. However, the gas and total
gravitating mass are determined from the X-ray data alone, and so here we
are able to improve our statistics by using data from the full sample, 
rather than being limited to those systems for which we have optical data.

\subsection{Dark matter distribution}
\label{ssec:dm_radial}
\begin{figure}
\hspace{-0.3cm}
\includegraphics[angle=270,width=9cm]{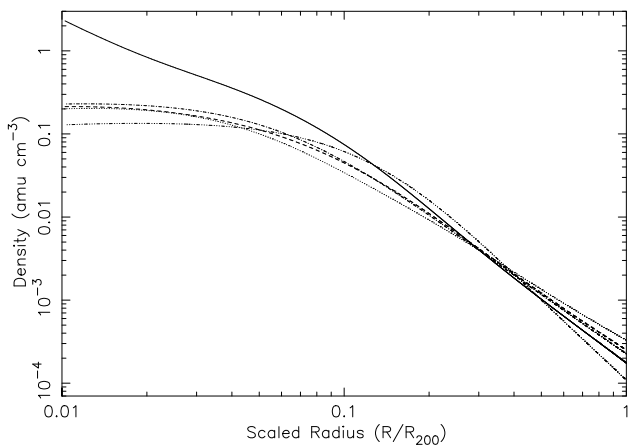}
\caption{ \label{fig:dm_profiles}
  The variation of dark matter density with scaled radius, for the
  \textbf{optical sample}, grouped by system temperature. The solid line
  represents the coolest systems (0.3--2.0\keV), increasing in temperature
  through dashed (2.0--2.9\keV), dotted (2.9--4.6\keV), dot-dashed
  (4.6--8.5\keV) and finally dot-dot-dot-dashed (8.5--17\keV).}
\end{figure}

The spatial variation of the density of dark matter can be seen in
Fig.~\ref{fig:dm_profiles}, for the optical sample. The similarity between
the hottest four temperature bands is quite close, with only the coolest
systems showing a deviation from the general trend, and then only within
$\sim$0.1\R200. The dark matter distribution is largely self-similar, as a
consequence of its insensitivity to the types of heating and/or cooling
processes which can influence baryonic material. Despite this, it is clear
that there is evidence of an enhanced central density in the core of the
average density profiles of the coolest groups. We return to this unusual
behaviour in more detail in section~\ref{dm_conc}.

\subsection{Total density}
\label{ssec:dtot}
\begin{figure}
\hspace{-0.3cm}
  \includegraphics[angle=270,width=9cm]{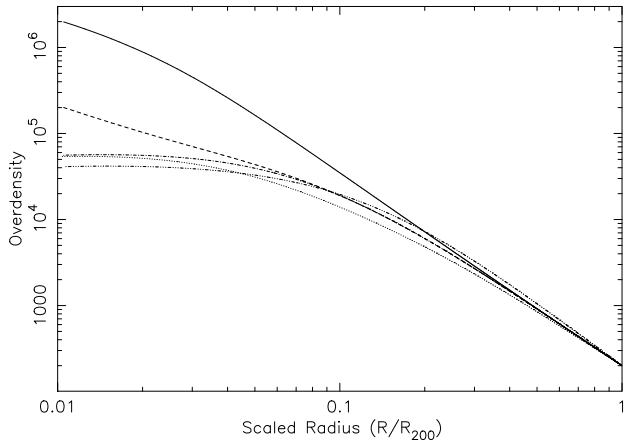}
\caption{ \label{fig:dover_profiles}
  Spatial variation of cumulative total density -- normalized to the
  critical density of the Universe -- with scaled radius, for the
  \textbf{full sample}, grouped by system temperature. The solid line
  represents the coolest systems (including the two galaxies)
  (0.3--1.3\keV), increasing in temperature through dashed (1.3--2.9\keV),
  dotted (2.9--4.6\keV), dot-dashed (4.6--8\keV) and finally
  dot-dot-dot-dashed (8--17\keV).}
\end{figure}

To provide some comparison with the density profiles of the separate mass
components we have plotted the profiles of integrated overdensity in
Fig.~\ref{fig:dover_profiles}, \ie\ the mean density within a given radius,
normalized to the critical density of the Universe. Since the gravitating
mass profile is determined from the X-ray data,
Fig.~\ref{fig:dover_profiles} incorporates our whole sample, including the
two early-type galaxies. Correspondingly, we have a much better coverage 
of the low mass end of the sample, and so have used a narrower temperature 
range for the coolest band (0.3--1.3\keV, instead of 0.3--2.0\keV) to 
provide greater sensitivity to any possible enhancements in density 
concentration at this mass scale.

The dark matter accounts for the majority of the gravitating mass, so it is
not surprising that the trend seen in Fig.~\ref{fig:dm_profiles} is
present in Fig.~\ref{fig:dover_profiles}. The difference between the dark
matter and total mass density seen in the second coolest bin can be
attributed to the slightly different temperature ranges in each case, used
for reasons explained previously.

Although the total density profiles in Fig.~\ref{fig:dover_profiles} are
essentially self-similar in the outer regions, there is still a reasonable
degree of scatter. We have compared \R200\ with two different radii of
overdensity, for each object in our X-ray sample: we have selected
\rmsub{R}{2500} and \rmsub{R}{500}, which are often used in the literature
\citep[\eg][]{all01,fin01}, since X-ray haloes are more readily traceable
out to these radii than to \R200. We find that, on average, an overdensity
of 2500 and 500 correspond to 31 and 66 per cent of \R200, respectively,
with corresponding standard deviations across our sample of 5 and 4 per
cent.

\subsection{Gas distribution}

\begin{figure}
\hspace{-0.3cm}
\includegraphics[angle=270,width=9cm]{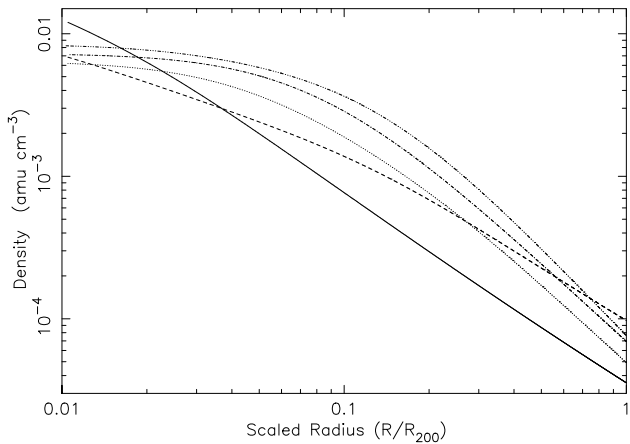}
\caption{ \label{fig:gas_profiles}
  The variation of gas density with scaled radius, for the
  \textbf{full sample}, grouped by system temperature. The solid line
  represents the coolest systems (including the two galaxies)
  (0.3--1.3\keV), increasing in temperature through dashed (1.3--2.9\keV),
  dotted (2.9--4.6\keV), dot-dashed (4.6--8\keV) and finally
  dot-dot-dot-dashed (8--17\keV).}
\end{figure}

Fig.~\ref{fig:gas_profiles} shows the gas density profiles for the whole
X-ray sample, averaged according to their mean X-ray temperature. It is
very clear that there is a great deal of variation in the five different
curves. Even for the three profiles representing clusters hotter than
2.9\keV\ there is a variation in the normalization of the lines, although
the shape of the profiles is very similar.

The two coolest temperature bands exhibit very different behaviour, having
the lowest gas density within 0.3\R200\ and showing no evidence of the
central core seen in the hotter clusters. The normalization of the cool
group lines is consistent with the trend towards a lowering of gas density
compared to the clusters. This behaviour mirrors the trend seen in the gas
fraction in \citetalias{san03}, and provides strong evidence of a deviation
from self-similar scaling of the IGM both within as well as between haloes
of different masses.

\subsection{Central density concentration}
\label{dm_conc}

\begin{figure*}
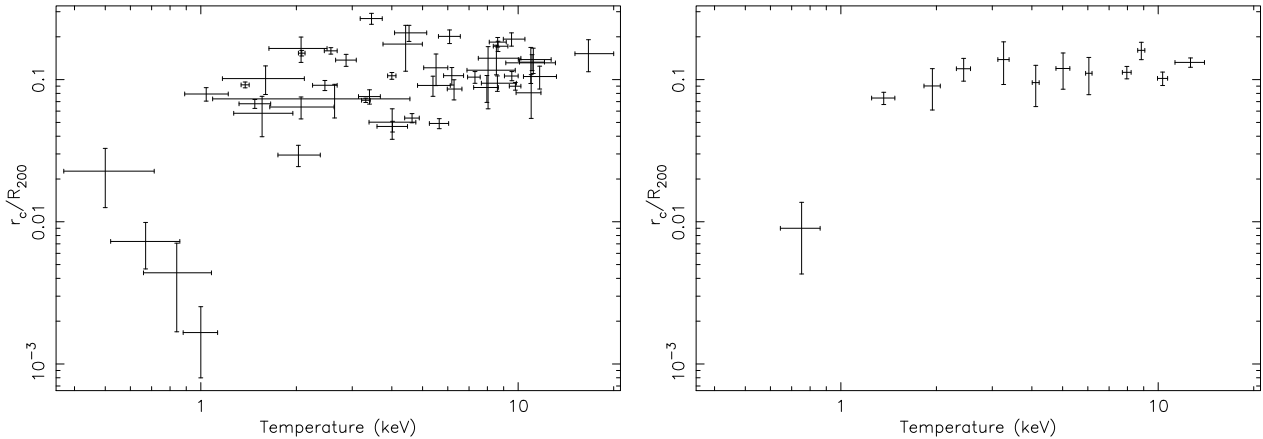

  \centering
  \subfigure{
    \includegraphics[angle=270,width=9.0cm]{fig9a.eps}}
  \hspace{-0.7cm}
  \subfigure{
    \includegraphics[angle=270,width=9.0cm]{fig9b.eps}}
  \vspace{-0.3cm}
  \caption{The ratio of \rmsub{r}{c} to \R200\ as a function of 
    temperature. The left panel shows all the individual points; the right
    panel shows the data grouped to a minimum of four points per bin. The 
    axes in each plot have been scaled identically. The 18 systems with 
    potentially unreliable core radii have been excluded (see text for 
    details).}
  \label{fig:rc_r200_kT}   
\end{figure*}

Although the underlying gravitational potential in virialized systems is
expected to be self similar, simulations indicate that the concentration of
the dark matter should vary slightly with mass \citep{nav97}. It was shown
in Figs.~\ref{fig:dm_profiles} and \ref{fig:dover_profiles} that the
coolest systems in our sample do indeed appear to be more centrally
concentrated -- an effect which is most pronounced in the central $\sim$10 
per cent of \R200.

To understand what is driving this behaviour, we have examined the scaling
properties of the constant density core in the gas distribution, \rc, as
measured from the X-ray data in \citetalias{san03}. The left panel of
Fig.~\ref{fig:rc_r200_kT} shows \rc\ as a fraction of \R200\ plotted
against system temperature. For the hottest clusters the points scatter
about a self-similar mean value of roughly 10 per cent. As the temperature
decreases, however, the scatter increases and $\la$1\keV\ there is a very
sharp drop. This trend is in excellent agreement with the predictions of
the galaxy formation-regulated gas evolution model of \citet{bry00}, as
studied in detail by \citet{wuX02b}: there is a remarkable similarity
between Fig.~\ref{fig:rc_r200_kT} and the model and data points in fig.~8
from the latter paper.

To minimize the possibility of systematic bias in this result, we have
identified and excluded from Fig.~\ref{fig:rc_r200_kT} those systems for
which the core radius may be unreliable, owing to the presence of a central
cooling emission excess. In 17 cases, \rc\ was measured to be
\emph{smaller} than the either the size of the radius used to excise the
cooling flow or the radius within which a cooling flow component was
fitted. In addition, the cluster Abell~2218 was excluded, since it was
necessary to fix its core radius at the best-fitting value in order to
stabilize the fitting during the calculation of parameter errors
\citepalias{san03}, leaving a total of 48 systems plotted in
Fig.~\ref{fig:rc_r200_kT}. To suppress the scatter in the relation, we have
grouped the points together to a minimum of four points per bin, giving a
total of 12 bins. The errors on each point are determined from the scatter
in the X and Y directions.  This plot is shown in the right panel of
Fig.~\ref{fig:rc_r200_kT}. The underlying trend in the data is much clearer
-- it can be seen that the coolest bin is at least 10$\sigma$ lower than the
flat relation established by the six hottest bins.

It is difficult to explain this behaviour in terms of a fitting bias: had
these coolest systems exhibited X-ray core radii consistent with the
cluster trend, \ie\ $\ga$ 3 times larger than observed, they would very
easily have been detected. One possibility is that the values of \R200\ are
anomalously large. To test this we have plotted \R200\ against temperature
in Fig.~\ref{fig:R200_kT}. The slope of this relation is somewhat steeper
than the self-similar prediction of 0.5, but there is clearly no evidence
of a trend towards an unusually large \R200\ in the cooler systems, which
could account for the discontinuity observed in \rmsub{r}{c}/\R200. The
most clearly discrepant point at the low-mass end is the S0 galaxy NGC~1553
(the coolest system in our sample), but it appears to have a value of
\R200\ which is \emph{lower} than its temperature would suggest. However,
this object has unusual properties, which point to an anomalously high
temperature, which may have been boosted by energy injection from stellar
winds (see \citetalias{san03}).

\subsection{Star formation efficiency and gas loss from haloes}
\label{ssec:SFE}

\begin{figure*}
  \centering
  \subfigure{
    \includegraphics[angle=270,width=9.0cm]{fig10a.eps}}
  \hspace{-0.7cm}
  \subfigure{
    \includegraphics[angle=270,width=9.0cm]{fig10b.eps}}
  \vspace{-0.3cm}
  \subfigure{
    \includegraphics[angle=270,width=9.0cm]{fig10c.eps}}
  \hspace{-0.7cm}
  \subfigure{
    \includegraphics[angle=270,width=9.0cm]{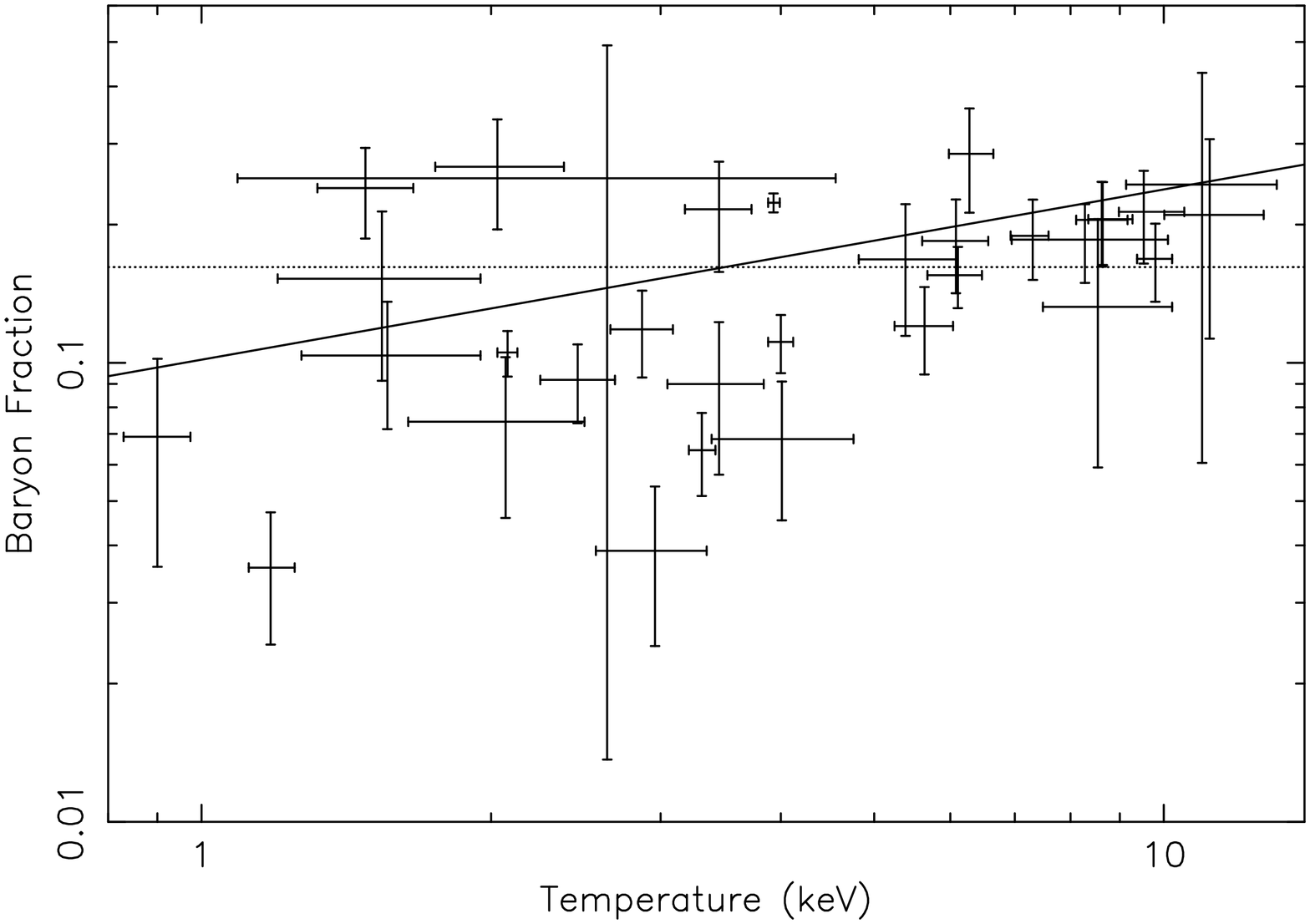}}
  \vspace{-0.3cm}
  \caption{\textit{Top left panel}: The fraction of baryons in the 
  form of stars, within \R200, as a function of system temperature, for the 
  optical sample. \textit{Top right}: The ratio of stellar to 
  dark matter mass within \R200, as a function of system temperature, for 
  the optical sample. \textit{Bottom left}: The ratio of gas to dark 
  matter mass within \R200, as a function of system temperature, for the 
  optical sample. \textit{Bottom right}: Baryon fraction within \R200\ 
  as a function of system temperature, for the optical sample. The solid 
  line represents the best-fitting power law, which has a logarithmic slope 
  of $0.37\pm0.15$.  The dotted line marks the median value of 0.161.}
  \label{fig:mass_ratio_mosaic}   
\end{figure*}

\begin{figure}
\hspace{-0.3cm}
\includegraphics[angle=270,width=9cm]{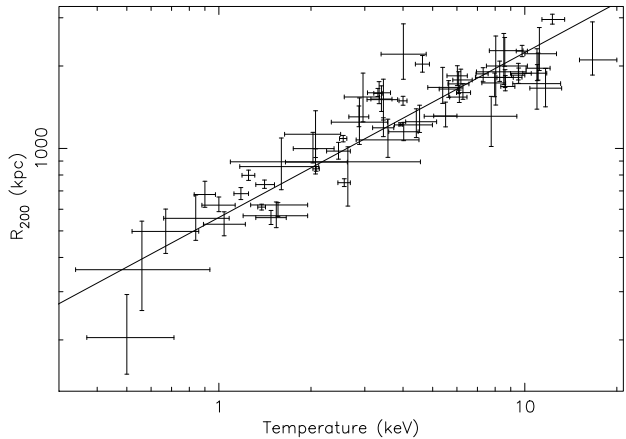}
\caption{ \label{fig:R200_kT}
  The variation of \R200\ as a function of temperature.  The solid line
  depicts the best fitting linear relation, with a slope of $0.60 \pm
  0.03$.}
\end{figure}

To address the issue of star formation efficiency, we have examined the
star-to-baryon ratio as a measure of the effectiveness with which gas has
been converted into stellar material. The upper left panel of
Fig.~\ref{fig:mass_ratio_mosaic} shows this quantity plotted against system
temperature for the optical sample. It can be seen that there is some
evidence of a negative correlation, which is significant at the $3.1\sigma$
level. However, this trend is potentially misleading, since it may be due
at least in part to the variation in gas mass with temperature seen
above. This means that some allowance must be made for the possibility that
gas may be lost to the system beyond the virial radius, \ie\ the gas mass
may be reduced without any corresponding increase in the stellar mass.

A better discriminator of star formation efficiency is obtained if the
stellar mass is normalized to the mass of dark matter, since this component
is largely immune to bias from non-gravitational processes. We have plotted
this ratio against X-ray temperature in the upper right panel of
Fig.~\ref{fig:mass_ratio_mosaic}. It is clear that the trend seen in upper
left panel of Fig.~\ref{fig:mass_ratio_mosaic} has largely vanished,
leaving only 0.8$\sigma$ evidence for a systematic variation in star
formation efficiency with halo temperature.

To identify the cause of the trend in the top left panel of
Fig.~\ref{fig:mass_ratio_mosaic}, we have also plotted the ratio of gas to
dark matter mass, shown in the lower left panel of
Fig.~\ref{fig:mass_ratio_mosaic}. A positive correlation between this ratio
and the X-ray temperature is significant at the $3.3\sigma$ level.  This
confirms that the variation in gas fraction with temperature does indeed
fully account for the trend found in the stellar-to-baryon ratio.

\subsection{Baryon fraction and constraints on $\rmsub{\Omega}{m}$}
Using our measured stellar and gas mass data, we are able to determine the
fraction of mass in baryons, \fbary, for our optical sample, which we show
plotted against X-ray temperature in the lower right panel of
Fig.~\ref{fig:mass_ratio_mosaic}. There is evidence of a modest trend in
the data ($2.5\sigma$ significance, using Kendall's K statistic -- a 
non-parametric correlation test). The best-fitting power law is
represented by the solid line, and is given by
$\log{\fbary}=(0.37\pm0.15)\log{kT}-(0.99\pm0.10)$; the median value of
0.161 is indicated by the dotted line. 

If virialized systems constitute a fair sample of the baryon content of the
Universe, we can use this median value to constrain the total mass density,
$\rmsub{\Omega}{m} = \Omega_b/\fbary$.  Assuming a baryon density of
$\Omega_b$ = 0.044 $\pm$ 0.004 \h70$^{-2}$, from measurements of the power
spectrum of the cosmic microwave background, using \WMAP\ \citep{bennett03}, 
we infer a value of $\rmsub{\Omega}{m} = 0.27\,\h70^{-1}$. This agrees well 
with the results of \citet{hra00}. 

If we improve our statistics by combining our mean stellar fraction, of
$(0.034 \pm 0.006)$, with our mean gas fraction for the full X-ray sample
\citepalias{san03}, of $(0.134 \pm 0.01)\,\h70^{-3/2}$, \fbary\ increases 
to 0.168 and $\rmsub{\Omega}{m}$ drops to 0.26. Assuming an unbiased 
measurement of the stellar and gas fraction, this represents a upper limit 
on $\rmsub{\Omega}{m}$, since any baryonic component to the dark matter 
would increase \fbary. Both these values are in good  agreement with the 
latest measurements of $\rmsub{\Omega}{m} = 0.27\pm0.04$ from \WMAP\ 
\citep{bennett03}.

This limit on $\rmsub{\Omega}{m}$ is slightly lower than the value of
$0.30^{+0.04}_{-0.03}$ inferred by \citet{all02}, based on a Chandra
analysis of six massive lensing clusters. The discrepancy arises partly
from a difference in gas fraction and partly from \citeauthor{all02}'s
choice of stellar to gas ratio. They use a value of $0.19\,h^{0.5}$ for the
latter quantity \citep{fuk98b,whi93}, which is rather less than our
measured value of $0.30\pm0.04$. However, the latter value of 0.30 is
heavily biased by a handful of groups, which have a particularly high
fraction of stellar baryons (the uppermost points in the top left panel of
Fig.~\ref{fig:mass_ratio_mosaic}); the median stellar to gas ratio for our
sample is 0.21, in much better agreement with existing measurements, which
are based on more massive clusters.

Moreover, the \citeauthor{all02} mean measured gas fraction is
$0.113\pm0.005$, compared to our mean of $0.134\pm0.01$ for the whole
X-ray sample \citepalias{san03}. The cause of the discrepancy in both
cases is the scaling behaviour of the gas fraction -- we find evidence of
a rise in gas fraction with radius, whereas \citeauthor{all02} report a
substantially flat \fgas\ profile in a number of their clusters, albeit
restricted to a radius of overdensity of \rmsub{R}{2500}. By extrapolating
a constant value out to \R200, no allowance is made for an increase in
\fgas, as is predicted by numerical simulations, even when the effects of
preheating and radiative cooling are absent \citep{eke98,fre99}.

Furthermore, we find evidence of a decrease in \fgas\ in cooler systems,
which has the effect of lowering our mean value. Given the possible impact
of non-gravitational heating on low mass systems, the gas fraction in
richer clusters is likely to be a better indicator of the universal value.
If we calculate an average gas fraction for our hottest clusters
($>$5\keV), we obtain an even higher value, of $0.17\pm0.01$. Combined with
our mean stellar fraction from above, this places a more stringent upper
limit on the mass density, of $\rmsub{\Omega}{m} \leq 0.22\,\h70^{-1}$,
which lies just below the WMAP 1$\sigma$ confidence interval.

\section{Discussion}
\label{sec:discuss}

\subsection{Implications for heating/cooling}
Our results confirm the systematic breaking of self-similarity in the IGM
that were observed in the gas fraction in \citetalias{san03}. Clearly the
thermal history of the hot gas has been altered by the influence of
non-gravitational physics. The most promising candidates are radiative
cooling \citep{bry00,mua01} and energy injection by heating
\citep[\eg][]{val99}.  Both mechanisms are able to account for the X-ray
observations: more efficient cooling in denser (smaller) haloes leads to a
depletion of gas in the inner regions; the energetically boosted IGM is
only weakly captured in the shallower potential wells of less massive
virialized systems, thus reducing their gas fraction.

The apparent self-similarity of the stellar distribution with respect to
the dark matter suggests that star formation has similar efficiency in both
groups and clusters. However, the behaviour of the gas to dark matter ratio
points to gas loss in the cooler systems, albeit subject to significant
extrapolation of the data.  Together, these results are more suggestive of
non-gravitational heating as the likely mechanism responsible for the
observed breaking of self-similarity in virialized systems, since this
would more naturally account for depletion of gas in cooler systems without
a corresponding enhanced stellar mass. The alternative, radiative cooling,
would only be able to reduce gas mass by associated star formation, unless
it is able to form baryonic dark matter -- for example, molecular clouds
\citep[\eg][]{pfe94}.

The cooling hypothesis \citep[\cf][]{kni97} has received increased
attention lately, particularly in relation to the observed entropy
properties of virialized systems -- indeed our own results
\citep[see][paper III]{pon03} are consistent with some of the predictions of
such models, when the effects of feedback from associated star formation
are allowed for \citep{voi01}.  However, the natural consequence of
increased gas cooling is enhanced star formation, unless a substantial
reservoir of baryons exists in the form of less luminous matter -- \eg\
molecular clouds \citep{edg01}. The stellar properties of our sample of
virialized systems appear consistent with the predictions of self-similar
scaling, and we find no evidence for a significant variation in star
formation efficiency across a wide range of halo masses. Therefore, we
conclude that heating must have had at least some role to play in
systematically modifying the properties of the IGM.

\subsection{Halo formation epoch}
From both the dark matter and total mass density profiles, it is clear that
there is an enhancement in the central concentration in the very
coolest systems in our sample. This behaviour is consistent with
hierarchical structure formation \citep[\eg][]{blu84}, in which the
smallest haloes collapse at the earliest epochs: the higher density of the 
Universe at this time results in a higher central density. Such a systematic 
variation in the epoch of formation of virialized systems has previously 
been observed \citep{sat00}.

Another possibility is that the greater ages for these systems allow more
time for accumulation of extra material on to their haloes, via accretion
\citep{sal98}. This process would have the effect of increasing the virial
radius, but without significantly modifying the characteristic turnover
radius in the gas density profile, \ie\ the X-ray core radius as a fraction
of \R200\ would shrink, as observed in section~\ref{dm_conc}. Although
small haloes are still able to form at the present epoch, there may be an
intrinsic bias towards observing older, more relaxed systems, which are
likely to be brighter -- indeed, we selected our X-ray sample principally
on the basis of a regular X-ray morphology. Thus we might expect to find
fewer examples of cool systems with large a core, compared to their virial
radius. This may account for the absence of points around the cluster trend
of $\rc/\R200 \approx 0.1$ below $\sim$1\keV\ in Fig.~\ref{fig:rc_r200_kT}.

The formation epoch of virialized haloes may play quite a significant role
in influencing their scaling properties. This is particular true for nearby
systems, where the redshift of observation, $\rmsub{z}{obs}$, is a poor and
systematically biased estimator of the redshift of formation,
$\rmsub{z}{f}$. The trend for less massive haloes to be older may account
for the observed steepening of the \MT\ relation \citepalias[see][]{san03},
in contrast to the apparently self-similar slope found in more distant,
massive clusters \citep{all01}, where $\rmsub{z}{obs}$ is a much
better measure of $\rmsub{z}{f}$. However, the simulations of
\citet{mat01b} indicate that this process has a negligible effect on the
temperatures of those clusters which have assembled 75 per cent of their
final mass by $z=0.6$.

\section{Conclusions}
We have conducted a detailed study of the mass composition of a large
sample of virialized systems. Using X-ray data for our whole sample of 66
objects, we have derived the gas and total gravitating mass profiles as a
function of radius. Optical data for a subsample of 32 groups and clusters
have allowed us to determine the stellar mass distribution, and thus infer
the dark matter density profile.

We have determined the deprojected luminosity distribution from unweighted
surface density fits to galaxy positions from the APM survey, combined with
aperture luminosity measurements taken from a small number of sources in
the literature. Using the galaxy density profile we are able to extrapolate
the light to our nominal virial radius of \R200\ (as determined from the
X-ray data) to yield the \emph{deprojected} optical luminosity, in the
\Bj\ photometric band.

The scaling properties of the total $B$ band optical luminosity within
\R200\ are consistent with self-similarity; $M\propto L^{1.08\pm 0.12}$ and
$L\propto T^{1.62\pm 0.14}$, for total mass and emission-weighted
temperature, respectively. Similarly, the mass-to-light ratio remains
essentially constant across the sample, showing no trend with X-ray
temperature ($\MLR \propto T^{-0.06\pm0.17}$); we find a logarithmic mean
\MLR\ of $233^{+32}_{-28}$\,\h70\ \MLRBjsun, in good agreement with other
measurements based on X-ray mass estimates. Our data therefore provide no
evidence for a significant increase in star formation rate in galaxy
groups, which is constant across most of the sample.

We find that the dark matter is the most centrally concentrated mass
component, followed by the galaxies, with the X-ray emitting IGM the most
extended component. Comparing the spatial distribution of these mass
components across two decades of halo mass, we find that the dark matter
and total mass density profiles of our sample are nearly self-similar, but
for a clear central excess in the coolest systems ($\la 1.5\keV$). We
attribute this enhancement in central density concentration to a sharp
decline in the size of the gas core radius, \rc\ (normalized to \R200)
amongst systems cooler than $\sim$1\keV. Surprisingly, we find that there
is very little variation in the shape and normalization of the stellar
density profile across the sample. However, the gas density clearly departs
from this trend: the IGM is significantly more extended and less dense in
smaller haloes.

We measure a mean stellar mass fraction of $(0.032 \pm 0.004)$ and a median
baryon fraction, for our optical sample, of $0.161\,\h70^{-3/2}$.  This
allows us to place an upper limit of the mass density of the Universe of
$\rmsub{\Omega}{m} \leq 0.27\,\h70^{-1}$, though it is possible that
somewhat lower values are indicated by our data: if we restrict our
analysis to clusters hotter than 5\keV\ we deduce $\rmsub{\Omega}{m} \leq
0.22\,\h70^{-1}$. This change is a consequence of a trend in \fgas\ with
temperature, resulting from the influence of non-gravitational processes on
the intracluster gas. Our results favour energy injection by
non-gravitational heating as a contributory mechanism for explaining the
observed breaking of self-similarity in the IGM, since radiative cooling
alone would lead to an significant increase in star formation efficiency in
groups, in contrast to our findings.

As improved X-ray mass estimates are obtained, with \XMM\ and \Chandra, and
the technology of wide-field optical CCD photometry continues to advance,
the prospects for very accurate determination of the mass-to-light ratio
and baryon content in virialized systems look good.

\section*{Acknowledgments}
We are grateful to Alexis Finoguenov, Ed Lloyd-Davies and Maxim Markevitch
for providing the X-ray data and contributing to the original
analysis. AJRS thanks Alastair Edge and Somak Raychaudhury for helpful
comments and suggestions and Yen-Ting Lin for useful discussions. We are
grateful to the referee for suggesting several improvements to the paper. AJRS
acknowledges financial support from the University of Birmingham. This work
has made use of the Starlink facilities at Birmingham, the Automatic Plate
Measuring (APM) machine catalogue at Cambridge and the NASA/IPAC
Extragalactic Database (NED).\\

\textit{Astro-ph version only:
Many thanks to Vince Eke for spotting a mistake in the photometric conversion
from the $B$ to \Bj\ bands (equation~\ref{eqn:B2Bj}), which affected 3 cluster
luminosities.}

\bibliography{$AJRS_LATEX/ajrs_bibtex} 
\label{lastpage}

\end{document}

%% file: table1.tex
%


%
%

\begin{table*}
\begin{tabular}{l*{9}{c}}
\hline
Name & RA & Dec. & $z$ & T$^{a}$ & \R200\ & $r_s^{b}$ & \LBj\ & \MLR\ & Reference$^{c}$ \\[0.5ex]
  & (J2000) & (J2000) &  & (keV) & (arcmin) & (arcmin) & $(10^{11} \rmsub{L}{B,j,\sun})$ & \MLRBjsun\  & \\[0.5ex]
\hline\hline\\[-2ex] 
NGC 4325    & 185.825 & 10.622 & 0.0252 & 0.90 & 22.4 & 2.6 & $1.22_{-0.6}^{+0.6}$ & $293_{-160}^{+160}$  & G02 (APS) \\
NGC 5846    & 226.385 & 1.696 & 0.0058 & 1.18 & 95.5 & 39.3 & $1.80_{-0.95}^{+0.95}$ & $204_{-107}^{+107}$  & H02 \\
HCG 62      & 193.284 & -9.224 & 0.0137 & 1.48 & 33.5 & 12.7$^{\star}$ & $2.59_{-1.3}^{+1.3}$ & $78_{-40}^{+40}$  & G02 (APS) \\
NGC 5129    & 201.150 & 13.928 & 0.0233 & 1.54 & 20.5 & 7.8 & $2.56_{-1.3}^{+1.3}$ & $82_{-48}^{+48}$  & G02 (APS) \\
NGC 2563    & 125.102 & 21.096 & 0.0163 & 1.61 & 31.4 & 12.1 & $1.76_{-0.9}^{+0.9}$ & $161_{-82}^{+82}$  & H02 \\
Abell 262   & 28.191 & 36.157 & 0.0163 & 2.03 & 50.4 & 47.0$^{\star}$ & $9.71_{-4.9}^{+4.9}$ & $118_{-65}^{+65}$  & G02 (APS) \\
Abell 194   & 21.460 & -1.365 & 0.0180 & 2.07 & 52.2 & 20.4 & $5.16_{-2.6}^{+2.6}$ & $320_{-200}^{+200}$  & G02 (COSMOS) \\
MKW 4       & 180.990 & 1.888 & 0.0200 & 2.08 & 35.5 & 40.3 & $1.12_{-0.6}^{+0.6}$ & $613_{-310}^{+310}$  & G02 (APS) \\
MKW 4S      & 181.647 & 28.180 & 0.0283 & 2.46 & 28.9 & 11.9 & $3.37_{-1.2}^{+1.2}$ & $320_{-130}^{+130}$  & H00 \\
NGC 6338    & 258.825 & 57.400 & 0.0282 & 2.64 & 26.4 & 5.9 & $14.0_{-7.0}^{+7.0}$ & $59_{-50}^{+60}$  & G02 (APS) \\
Abell 539   & 79.134 & 6.442 & 0.0288 & 2.87 & 38.0 & 5.9 & $6.17_{-3.1}^{+3.1}$ & $415_{-200}^{+200}$  & G00 \\
AWM 4       & 241.238 & 23.946 & 0.0318 & 2.96 & 40.2 & 11.5 & $4.75_{-2.4}^{+2.4}$ & $886_{-540}^{+540}$  & G02 (APS) \\
Abell 1060  & 159.169 & -27.521 & 0.0124 & 3.31 & 104.8 & 15.0$^{\star}$ & $18.9_{-9.5}^{+9.5}$ & $244_{-120}^{+120}$  & G02 (COSMOS) \\
Abell 2634  & 354.615 & 27.022 & 0.0309 & 3.45 & 32.1 & 12.0 & $25.4_{-12.6}^{+12.8}$ & $76_{-40}^{+40}$  & G02 (APS) \\
Abell 2052  & 229.176 & 7.002 & 0.0353 & 3.45 & 36.6 & 2.3 & $6.70_{-1.7}^{+1.7}$ & $589_{-240}^{+240}$  & H00 \\
Abell 2199  & 247.165 & 39.550 & 0.0299 & 3.93 & 34.3 & 31.2$^{\star}$ & $9.98_{-2.4}^{+3.5}$ & $211_{-60}^{+60}$  & H00 \\
Abell 2063  & 230.757 & 8.580 & 0.0355 & 4.00 & 35.7 & 2.3 & $7.08_{-1.6}^{+1.7}$ & $541_{-140}^{+140}$  & H00 \\
AWM 7       & 43.634 & 41.586 & 0.0172 & 4.02 & 105.2 & 4.2$^{\star}$ & $7.68_{-3.8}^{+3.8}$ & $1614_{-970}^{+970}$  & G00 \\
Abell 3391  & 96.608 & -53.678 & 0.0536 & 5.39 & 26.9 & 4.3 & $25.1_{-12.6}^{+12.6}$ & $213_{-120}^{+120}$  & G02 (COSMOS) \\
Abell 2670  & 358.564 & -10.408 & 0.0759 & 5.64 & 19.6 & 5.9 & $20.7_{-10.4}^{+10.3}$ & $249_{-130}^{+130}$  & G02 (COSMOS) \\
Abell 119   & 14.054 & -1.235 & 0.0444 & 6.08 & 33.5 & 5.5 & $26.9_{-13.4}^{+13.6}$ & $218_{-120}^{+120}$  & G02 (COSMOS) \\
Abell 496   & 68.397 & -13.246 & 0.0331 & 6.11 & 39.2 & 20.0$^{\star}$ & $13.1_{-6.5}^{+6.6}$ & $321_{-170}^{+170}$  & G02 (COSMOS) \\
Abell 3558  & 201.991 & -31.488 & 0.0477 & 6.28 & 28.7 & 18.2 & $47.7_{-23.8}^{+23.8}$ & $99_{-50}^{+50}$  & G02 (COSMOS) \\
Abell 3571  & 206.867 & -32.854 & 0.0397 & 7.31 & 39.8 & 17.9 & $46.8_{-23.4}^{+23.4}$ & $161_{-80}^{+80}$  & G02 (COSMOS) \\
Abell 2218  & 248.970 & 66.214 & 0.1710 & 8.28 & 11.3 & 10.4$^{\star}$ & $47.5_{-6.0}^{+5.9}$ & $167_{-38}^{+38}$  & S96 \\
Abell 1795  & 207.218 & 26.598 & 0.0622 & 8.54 & 28.3 & 6.4$^{\star}$ & $27.9_{-14.0}^{+13.9}$ & $330_{-240}^{+240}$  & G02 (COSMOS) \\
Abell 2256  & 256.010 & 78.632 & 0.0581 & 8.62 & 27.3 & 7.3 & $39.2_{-19.6}^{+19.6}$ & $175_{-90}^{+90}$  & G00 \\
Abell 85    & 10.453 & -9.318 & 0.0521 & 8.64 & 28.1 & 11.6 & $23.9_{-12.0}^{+12.1}$ & $229_{-120}^{+120}$  & G02 (COSMOS) \\
Abell 3266  & 67.856 & -61.417 & 0.0545 & 9.53 & 29.7 & 12.0 & $29.3_{-14.7}^{+14.7}$ & $261_{-140}^{+140}$  & G02 (COSMOS) \\
Abell 2029  & 227.729 & 5.720 & 0.0766 & 9.80 & 26.4 & 4.6 & $86.3_{-43.1}^{+43.0}$ & $155_{-80}^{+80}$  & G02 (APS) \\
Abell 478   & 63.359 & 10.466 & 0.0882 & 10.95 & 17.8 & 12.8$^{\star}$ & $28.1_{-7.3}^{+7.3}$ & $210_{-160}^{+160}$  & H00 \\
Abell 2142  & 239.592 & 27.233 & 0.0894 & 11.16 & 22.7 & 8.7 & $50.0_{-25.0}^{+25.0}$ & $250_{-170}^{+170}$  & G02 (APS) \\
\hline
\end{tabular}
\caption{
 Some basic properties of the 32 objects in the optical sample, listed in order of 
 increasing temperature. Positions and redshifts are taken from
 \citet{ebe96,ebe98,pon96} and NED. Columns 7--9 are data as determined
 in this work, except for those values of $r_s$ marked with a $^{\star}$, which 
 were taken from \citet{llo01_thesis}. Note: values are for $H_{0}=70$\kmpspMpc. 
 All errors are 68\% confidence.
$^{a}$The cooling-flow corrected, emission-weighted temperature 
  of the system within 0.3\R200, as determined in \citetalias{san03}.
$^{b}$The NFW scale radius of the galaxy distribution (see equation~\ref{eqn:NFW}).
$^{c}$Reference for the luminosity data, G02 = \citet{gir02}; G00 = \citet{gir00};
  H00 = \citet{hra00}; H02 = \citet{hel03}; S96 = \citet{squ96}.
        }
\label{tab:sample}
\end{table*}